\documentclass[twocolumn,aps,prx,superscriptaddress]{revtex4}
\usepackage[latin9]{inputenc}
\usepackage{amsmath}
\usepackage{amssymb}
\usepackage{graphicx}

\makeatletter
\@ifundefined{textcolor}{}
{%
 \definecolor{BLACK}{gray}{0}
 \definecolor{WHITE}{gray}{1}
 \definecolor{RED}{rgb}{1,0,0}
 \definecolor{GREEN}{rgb}{0,1,0}
 \definecolor{BLUE}{rgb}{0,0,1}
 \definecolor{CYAN}{cmyk}{1,0,0,0}
 \definecolor{MAGENTA}{cmyk}{0,1,0,0}
 \definecolor{YELLOW}{cmyk}{0,0,1,0}
}

\usepackage{epsfig}
\usepackage{xcolor}
\usepackage{tikz}

\makeatother

\begin{document}

\title{Larkin-Ovchinikov superfluidity in time-reversal symmetric bilayer Fermi gases} 

\author{Qing Sun}
\address{Department of Physics, Capital Normal University, Beijing 100048,
China}

\author{Liang-Liang Wang}
\address{Institute for Natural Sciences, Westlake Institute for Advanced Study,
Westlake University, Hangzhou, Zhejiang Province, China}
\address{Westlake University, Hangzhou, Zhejiang Province, China}

\author{Xiong-Jun Liu}
\email{xiongjunliu@pku.edu.cn}
\address{International Center for Quantum Materials, School of Physics, Peking University, Beijing 100871, China}
\address{Collaborative Innovation Center of Quantum Matter, Beijing 100871, China}

\author{G. Juzeli\={u}nas}
\email{gediminas.juzeliunas@tfai.vu.lt}
\address{Institute of Theoretical Physics and Astronomy, Vilnius University,
 Saul\.etekio 3, LT-10257 Vilnius, Lithuania}

\author{An-Chun Ji}
\email{andrewjee@sina.com}
\address{Department of Physics, Capital Normal University, Beijing 100048,
China}

\date{\today }
\begin{abstract}
Larkin-Ovchinnikov (LO)  state which combines the superfluidity and spatial periodicity of pairing order parameter and exhibits the supersolid properties has been attracting intense attention in both condensed matter physics and ultracold atoms. Conventionally, realization of LO state from an intrinsic $s$-wave interacting system necessitates to break the time-reversal (TR) and sometimes spatial-inversion (SI) symmetries. Here we report a novel prediction that the LO state can be
realized in a TR and SI symmetric system representing a bilayer
Fermi gas subjected to a laser-assisted interlayer tunneling. We
show that the intralayer $s$-wave atomic interaction acts
effectively like a $p$-wave interaction in the pseudospin space.
This provides distinctive pairing effects in the present system with pseudspin
spin-orbit coupling, and leads to a spontaneous density-modulation of the pairing order predicted in a
very broad parameter regime. Unlike the conventional schemes, our
results do not rely on the spin imbalance or external Zeeman fields,
showing a highly feasible way to observe the long-sought-after LO
superfluid phase using the laser-assisted bilayer Fermi gases.
\end{abstract}
\maketitle

\section{introduction}
The Fulde-Ferrell-Larkin-Ovchinikov (FFLO) state with
finite-momentum pairing \cite{Fulde1964,Larkin1964} is an
exotic phase hosting many novel physical phenomena in condensed
matter physics and nuclear physics
\cite{Casalbuoni2004,Gubbels2012,Anglani2014}. In particular, the
predicted Larkin-Ovchinnikov (LO)  state exhibits a supersolidity,
which combines a superfluid order parameter and a unidirectional
Cooper-pair density wave, i.e. it simultaneously has the
off-diagonal long-range order and long-range density order. The two
orders are often mutually exclusive, and can lead to various exotic
low-energy modes \cite{Radzihovsky2010RPP} and macroscopic quantum
phenomena \cite{Boninsegni}. For bosons, a similar order, called stripe phase with
supersolid properties, has been recently observed in spin-orbit (SO)
coupled atomic Bose-Einstein condensates (BECs) \cite{Li2017Nature}.
However, although the LO state has been pursued for half a century,
the conclusive evidence of this exotic state remains
elusive so far for fermions.

Conventionally, realization of FFLO state needs to break the
time-reversal (TR) and sometimes spatial-inversion (SI) symmetries
for the $s$-wave interacting systems. For example, in the widely
studied spin imbalanced  Fermi gases
\cite{Zwierlein2006Science,Partridge2006Science,Sheehy2006PRL,Mizushima2005PRL,Hu2006PRA,Liao2010Nature},
the population imbalance is equivalent to a TR breaking Zeeman
field, which can induce mismatched Fermi surface and lead to the LO
state. In the SO coupled ultracold Fermi atom gas
\cite{Lin2011Nature,Zhang2012PRL,Wang2012PRL,Cheuk2012PRL,Qu2013PRA,Huang2016NP,Wu2016,Meng2016PRL,Galitski1,Goldma1,Zhai1,Zhang2018},
it was proposed that an in-plane Zeeman term can deform the
symmetric Fermi surface. This breaks the SI symmetry and induces the
Fulde-Ferrell (FF) superfluid whose pairing order has a single
non-zero momentum. The predicted FFLO states in the spin-orbit
coupling (SOC) systems
\cite{Chen2013PRL,Iskin2013PRA,Liu2013PRA,Wu2013PRL,Qu2013NC,Zhang2013NC,Zheng2016PRL,Dong2013PRA,Zheng2013PRA,LLWang,Poon}
are essentially the FF superfluids which preserve the translational
symmetry. The FFLO state was also proposed in the Weyl and Dirac
semimetals, where the broken TR and SI inversion symmetries may
favor the formation of finite-momentum pairing
orders~\cite{Chan2017-1,Chan2017-2}. However, since the broken TR
and SI symmetries generically suppress the $s$-wave superfluid
order, and near resonant Raman lasers induce heating problems in the
SOC systems~\cite{Zhang2018}, the FFLO phases predicted in the
literatures are hard to observe in experiment.

Here we  propose that the LO state with supersolid properties can be
realized in laser-assisted bilayer Fermi gases, which preserve both
the TR and SI symmetries. The layers play the role of a
pseudospin-$1/2$ system, and the laser-assisted interlayer tunneling
generates a 1D SOC in the pseudospin space ~\cite{Li2017Nature}. We
show that the $s$-wave interaction between the real spin-up and
spin-down states renders an effective $p$-wave interaction in the
pseudospin space, which essentially leads to a spontaneous LO
superfluid order in the presence of laser induced pseudospin SOC.
The present realization exhibits fundamental advantages over the
existing schemes in generating  the LO order. First, the pseudospin
SOC does not break TR symmetry, nor SI symmetry, for which the LO
phase can be obtained in a very broad parameter regime. Further,
generation of the pseudospin SOC does not apply near resonant
Raman couplings, so the present system can have a life time
 much longer than the cold atom systems with a real SOC. These
advantages enable the realization of LO order with the
currently available experimental techniques.

The paper is organized as follows: First in Sec. II, we formulate the model 
and analyze the symmetry of the system. Then based on the variational method 
and numerical simulations, we present the two-body problem and many-body phase 
diagram in Se. III. Finally in Sec. IV, we discuss some experimental-related issues and give
a brief summary.

\section{The model and symmetry analysis}
We consider a two-component Fermi
gas composed of atoms in two metastable internal (spin) states, for
example two hyperfine atomic ground states. The atoms are confined
in a state-independent double-well optical potential along the
$z$-axis providing a bilayer structure \cite{Li2016PRL,Sun2016}. An
asymmetry of the double well potential prevents a direct atomic
tunneling between the wells, and instead there is a laser-assisted
interlayer tunneling. The second-quantization Hamiltonian of this
bilayer system reads in the momentum space (see the Appendix for more details)
\begin{eqnarray}
\mathcal{H}\!\! & = &
\!\!\sum_{\mathbf{k}\gamma}\left[\sum_{j}\xi_{\mathbf{k}\gamma j}
\hat{\psi}_{\mathbf{k}\gamma j}^{\dag}\hat{\psi}_{\mathbf{k}\gamma
j}+\frac{J}{2}
(\hat{\psi}_{\mathbf{k}\gamma,1}^{\dag}\hat{\psi}_{\mathbf{k}\gamma,2}+h.c.)\right]\nonumber \\
 & + & \!\!\frac{U}{S}\sum_{\mathbf{kk^{\prime}q}j}\hat{\psi}_{\mathbf{k}\uparrow j}^{\dag}
 \hat{\psi}_{\mathbf{k^{\prime}}\downarrow j}^{\dag}\hat{\psi}_{\mathbf{k^{\prime}+q}\downarrow j}
 \hat{\psi}_{\mathbf{k-q}\uparrow j}\,,\label{eq:H}
\end{eqnarray}
with
\begin{equation}
\xi_{\mathbf{k}\gamma j}=\frac{1}{2}\left[(\mathbf{k}+(-1)^{j+1}\kappa\mathbf{e}_{x})^{2}+h(-1)^{j}\pm\delta\right]\,,
\label{eq:xi}
\end{equation}
where the upper and lower signs in Eq. (\ref{eq:xi}) correspond to
different atomic internal states labelled by
$\gamma=\uparrow,\downarrow$, and the atomic mass $m$ and $\hbar$
are set to the unity. Here $\hat{\psi}_{\mathbf{k}\gamma j}^{\dag}$
and $\hat{\psi}_{\mathbf{k}\gamma j}$ are Fermi operators for
creation and annihilation of an atom in the $j$th layer ($j=1,2$)
with the spin $\gamma$ and momentum $\mathbf{k}$ in the
$xy$ (layer) plane; $h$ and $\delta$ denote the energy mismatch
between the two layers and two internal states, respectively; $J$ is a
strength of the laser-assisted interlayer tunneling with
$2\kappa\mathbf{e}_{x}$ being an associated recoil momentum pointing
in the $x$-direction.

In writing Eqs. (\ref{eq:H})-(\ref{eq:xi}), we
have applied a gauge transformation
$\psi_{\gamma1}=e^{-i\kappa
x}\tilde{\psi}_{\gamma1}$ and $\psi_{\gamma2}=e^{i\kappa
x}\tilde{\psi}_{\gamma2}$,
where
$\tilde{\psi}_{\gamma j}$ is defined in the original (laboratory) frame (see Appendix).
Thus we are working in
the transformed basis involving the layer-dependent momentum shift
$(-1)^{j+1}\kappa\mathbf{e}_{x}$ featured in Eq.~(\ref{eq:xi}). This
gives a 1D pseudospin SOC in the $x$-direction for the
layer states, which is the same for both atomic internal states.
The gauge transformation makes the Hamiltonian given by Eq. (\ref{eq:H})
invariant under spatial translations in the $xy$ plane.

The atom-atom coupling in Eq. (\ref{eq:H}) is represented by the
contact interaction between the fermion atoms within individual
layers. In 2D systems  a bound state forms for  an arbitrarily small
attraction \cite{Randeria1989PRL}, and the contact interaction $U$
should be regularized by
$\frac{1}{U}=-\frac{1}{S}\sum_{\mathbf{k}}\frac{1}{E_{b}+2\epsilon_{\mathbf{k}}}$
 \cite{Randeria1}. Here $E_{b}$ is a binding energy of
the two-body bound state in the absence of  pseudospin SOC, $S$ is a
two-dimensional (2D) quantization volume and
$\epsilon_{\mathbf{k}}=\mathbf{k}^2 / 2$ is a 2D free particle
dispersion. For ultracold atoms $E_{b}$ can be tuned via the
Feshbach resonance technique.

The Hamiltonian ${\mathcal{H}}={\mathcal{H}}_0+\mathcal{H}_{\rm int}$ given by Eq.~ (\ref{eq:H}) can be recast in the basis of the
four-component spinor $\hat{\phi}=(\hat{\psi}_{\uparrow1}~\hat{\psi}_{\uparrow2}
~\hat{\psi}_{\downarrow1}~\hat{\psi}_{\downarrow2})^T$ defined in
the Hilbert space spanned by the tensor product of the spin and
layer states
 $(\uparrow,\downarrow)\otimes(1,2)$. Here ${\mathcal{H}}_0=\int
d^2\mathbf{r}\hat{\phi}^\dag(\mathbf{r}) h_0\hat{\phi}(\mathbf{r})$
is the single particle Hamitonian with
$h_0=\frac{\mathbf{k}^2+\kappa^2}{2} I_2\otimes
I_2+(\kappa\mathbf{k}\cdot\mathbf{e}_x-h)I_2\otimes\tau_z+\frac{J}{2}
I_2\otimes\tau_x+\delta\sigma_z\otimes I_2$, and $\mathcal{H}_{\rm
int} =\frac{U}{2S}\int
d^2\mathbf{r}\{[\hat{\phi}^\dag(\mathbf{r})I_2\otimes
I_2\hat{\phi}(\mathbf{r})]^2 +[\hat{\phi}^\dag(\mathbf{r})I_2\otimes
\tau_z\hat{\phi}(\mathbf{r})]^2\}$ is the interaction Hamitonian.
$I_2$ is the rank-2 unit matrix, $\sigma_z$ and $\tau_z$ are the
Pauli matrices defined in the real spin and pseudospin (layer)
spaces, respectively. For the real system one can set zero detunings
$h=\delta=0$ for convenience. In the presence of layer degree of
freedom, we can find that the system satisfies a TR symmetry denoted
as $T=i\sigma_y\tau_x K$, with $T{\mathcal{H}}T^{-1}={\mathcal{H}}$.
This symmetry operator is transformed back to the usual form
$T\rightarrow T_1=i\sigma_y K$ after applying a unitary rotation
$U=e^{-i\frac{\pi}{4}\tau_x}$ on the pseudospin space so that
$\tau_z\rightarrow\tau_y$, $\tau_y\rightarrow-\tau_z$, and ${\cal
H}\rightarrow\tilde{\mathcal{H}}$, which then satisfies
$T_1\tilde{\mathcal{H}}T_1^{-1}=\tilde{\mathcal{H}}$. This reflects
the fact that the pseudospin SOC does not induce transition between
real spins. Furthermore, the Hamiltonian (\ref{eq:H}) is invariant
also under the SI transformation involving the interchange of of the
two layers described by the operator $ \tau_x $ and the reversal of
the in-plane momentum: $\mathbf{k} \rightarrow -\mathbf{k}$. We note
that the $s$-wave interaction occurs between two fermions in the
real spin-up and spin-down states, respectively. As a consequence,
the existence of TR symmetry can greatly enhance the realization of
the LO state, as studied below.

\begin{figure}[t]
\centering \includegraphics[width=0.45\textwidth]{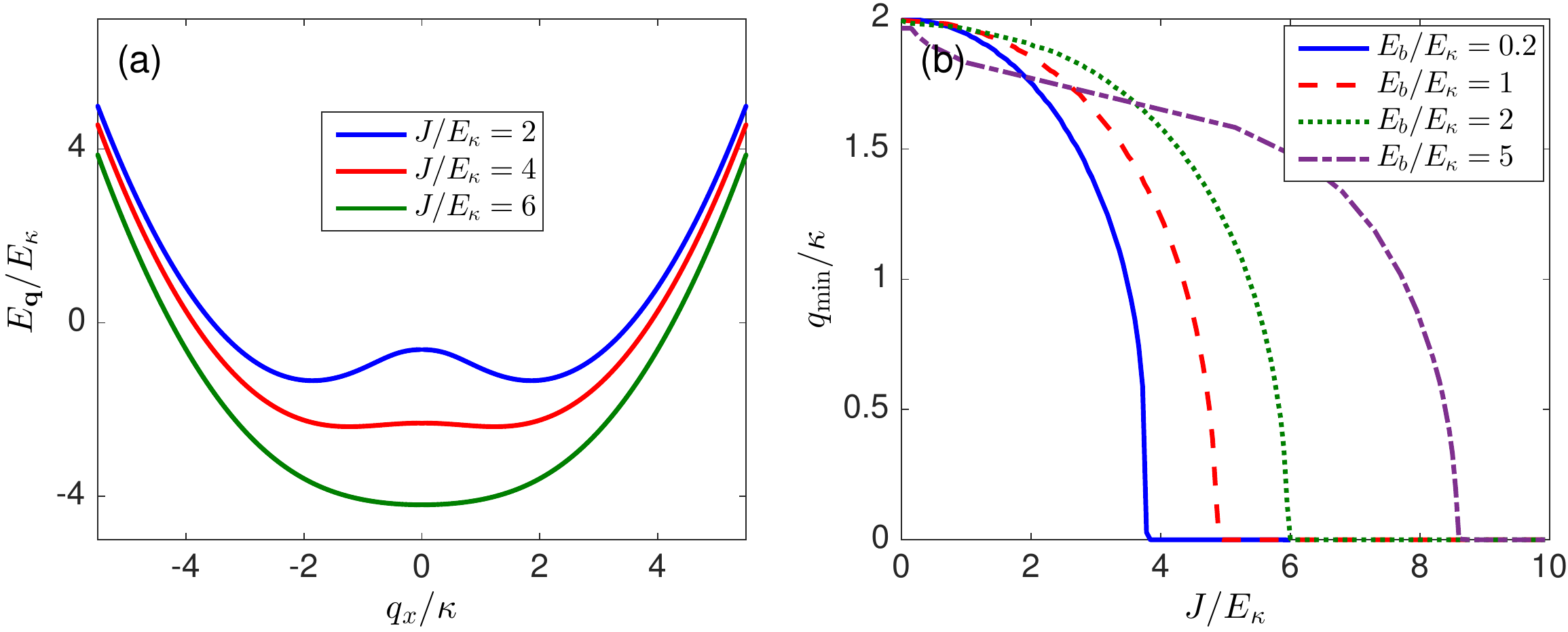}
\caption{{\bf Two-body bound state}. (a) The bound
state energy $E_{\mathbf{q}}$ as a function of COM momentum
$\mathbf{q}=q_{x}\hat{e}_x$ at $E_{b}/E_{\kappa}=1$, for
$J/E_{\kappa}=2$, 4, and 6. (b) The ground-state momentum of the
bound state  as a function of the tunneling strength $J$ for atomic
interaction $E_b/E_\kappa=0.2$, 1, 2, and 5. The energy and momentum
are measured in units of $E_{\kappa}=\kappa^{2}/2$ and $\kappa$,
respectively.} 
\label{molecular}
\end{figure}

\section{Results}
\subsection{Two-body Problem}
Since the pseudospin SOC does not mix the spin singlet and spin triplet, the wave function of
two-body bound states can be constructed in the singlet space as
\begin{eqnarray}
|\Phi\rangle_{\mathbf{q}}= \frac{1}{2}
\sum_{\mathbf{k}}\sum_{j,l=1}^{2}\phi_{\mathbf{k}\mathbf{q},jl}S_{\mathbf{k}
\mathbf{q},jl}^{\dag}\left|0\right\rangle ,\label{wave_function}
\end{eqnarray}
with
\begin{equation}
S_{\mathbf{k}\mathbf{q},jl}^{\dag}=\frac{1}{\sqrt{2}}[\psi_{j\uparrow}^{\dag}(\mathbf{Q}_{+})
\psi_{l\downarrow}^{\dag}(\mathbf{Q}_{-})-\psi_{j\downarrow}^{\dag}(\mathbf{Q}_{+})
\psi_{l\uparrow}^{\dag}(\mathbf{Q}_{-})]\label{eq:S^dagger}
\end{equation}
and $\mathbf{Q}_{\pm}=\mathbf{q}/2\pm\mathbf{k}$. Here
$S_{\mathbf{qk},jl}^{\dag}$ is a singlet operator for creating a
pair of atoms in the layers $j$ and $l$ with a center-of-mass (COM)
momentum $\mathbf{q}$ and a relative momentum $\mathbf{k}$, and
$\phi_{\mathbf{k}\mathbf{q},jl}$ is the corresponding amplitude. For
$j=l$ the atoms are paired in the same layer, whereas for $j\ne l$
the pairing takes place in different layers, the latter contribution
emerging due to the laser-assisted tunneling. Since
$S_{-\mathbf{k}\mathbf{q},jl}^{\dag}=S_{\mathbf{k}\mathbf{q},lj}^{\dag}$,
we take
$\phi_{-\mathbf{k}\mathbf{q},jl}=\phi_{\mathbf{k}\mathbf{q},lj}$ in
Eq. (\ref{wave_function}), in which the factor $1/2$ is to avoid a
double counting of the atomic pair states. Substituting
Eq.~(\ref{wave_function}) into the stationary Schr\"{o}dinger
equation
$\mathcal{H}|\Phi\rangle_{\mathbf{q}}=E_{\mathbf{q}}|\Phi\rangle_{\mathbf{q}}$,
one arrives at the following equation for the bound state energy
$E_{\mathbf{q}}$ (see Appendix for details).
\begin{eqnarray}
(\frac{U}{S}\sum_{\mathbf{k}}\frac{\alpha_{\mathbf{k}}}{\alpha_{\mathbf{k}}\gamma_{\mathbf{k}}
-\beta_{\mathbf{k}}^{2}}-1)(\frac{U}{S}\sum_{\mathbf{k}}\frac{\gamma_{\mathbf{k}}}{\alpha_{\mathbf{k}}
\gamma_{\mathbf{k}}-\beta_{\mathbf{k}}^{2}}-1)\nonumber \\
-(\frac{U}{S}\sum_{\mathbf{k}}\frac{\beta_{\mathbf{k}}}{\alpha_{\mathbf{k}}\gamma_{\mathbf{k}}
-\beta_{\mathbf{k}}^{2}})^{2}=0\,,\label{boundstate_energy}
\end{eqnarray}
where
$\alpha_{\mathbf{k}}=E_{\mathbf{q}}-(\mathbf{k}^{2}+\mathbf{q}^{2}/4+\kappa^{2})$,
$\beta_{\mathbf{k}}=q_{x}\kappa$ and
$\gamma_{\mathbf{k}}=\alpha_{\mathbf{k}}-\{\frac{J^{2}/2}
{\alpha_{\mathbf{k}}+2k_{x}\kappa}+\frac{J^{2}/2}{\alpha_{\mathbf{k}}-2k_{x}\kappa}\}$.
By solving Eq. (\ref{boundstate_energy}) we can determine
$E_{\mathbf{q}}$ as a function of the COM momentum.

Figure \ref{molecular}a illustrates that two degenerate minima are
formed at $\mathbf{q}_{0}=\pm q_{{\rm min}}\mathbf{e}_{x}$ in the
bound state energy spectrum $E_{\mathbf{q}}$. This is in sharp
contrast to the SOC for real spins, where the bound state always has
a zero COM momentum, or a single finite momentum if an in-plane
Zeeman field is applied to break the SI symmetry~\cite{Dong2013PRA}.
Furthermore, the interlayer singlet component
$P_{12}=\sum_{\mathbf{k}}|\phi_{\mathbf{k}\mathbf{q},12}|^{2}$ is
enhanced as the tunneling strength $J$ increases. Figure
\ref{molecular}b shows the numerical result of the minimal momentum
$q_{{\rm min}}=|\mathbf{q}_{0}|$ versus $J$ for the bound state. In
the strong tunneling limit where the tunneling strength exceeds a
critical value $J>J_c$, a bound state with  zero momentum is
eventually formed. On the other hand, we note that the critical
tunneling $J_c$ increases with $E_{b}$, showing that increasing
intralayer interaction enhances the formation of bound state with
finite-momentum.

\begin{figure}[t]
\centering \includegraphics[width=0.45\textwidth]{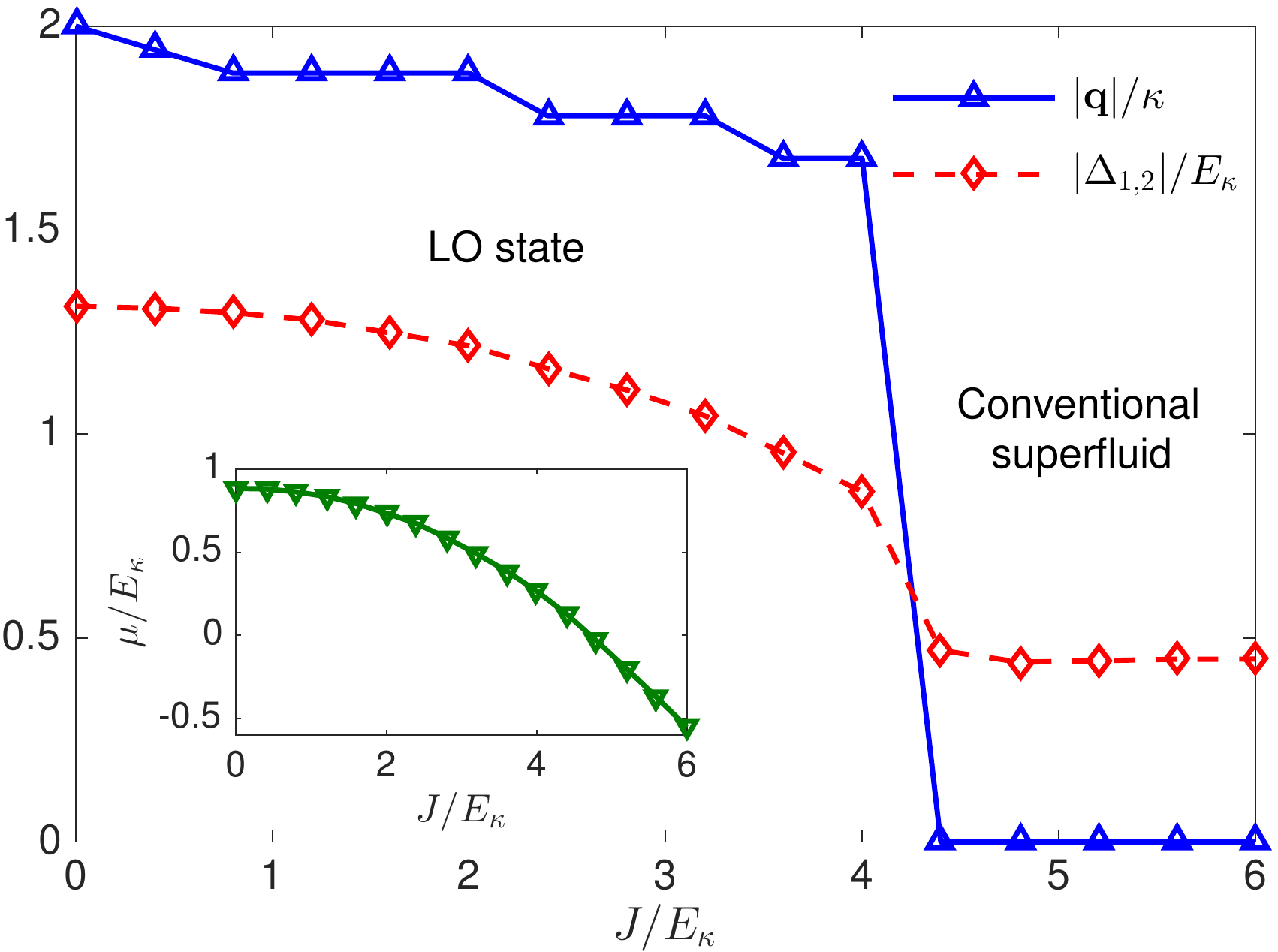}
\caption{{\bf Pairing orders of superfluid phases}.
Dependence of $|\Delta_{1,2}|$ (red dashed) and the pairing momentum
$|\mathbf{q}|$ (blue solid) on the tunneling strength $J$, where
$|\Delta_{1,2}|$ represents the maximum value of the order parameter
$|\Delta_{1,2}(x,y)|$ in the $xy$ plane. The inset shows the
evolution of the chemical potential $\mu$. In the simulations, we
have included a weak harmonic trap with frequency $\omega\ll
E_\kappa$, $R_{\rm
TF}\equiv\sqrt{2E_\kappa/m\omega^2}=\kappa/\omega$,
$E_{b}/E_{\kappa}=1$, and $E_{F}\sim E_{\kappa}$.}
\label{phase_diagram}
\end{figure}

\begin{figure}[t]
\centering \includegraphics[width=0.45\textwidth]{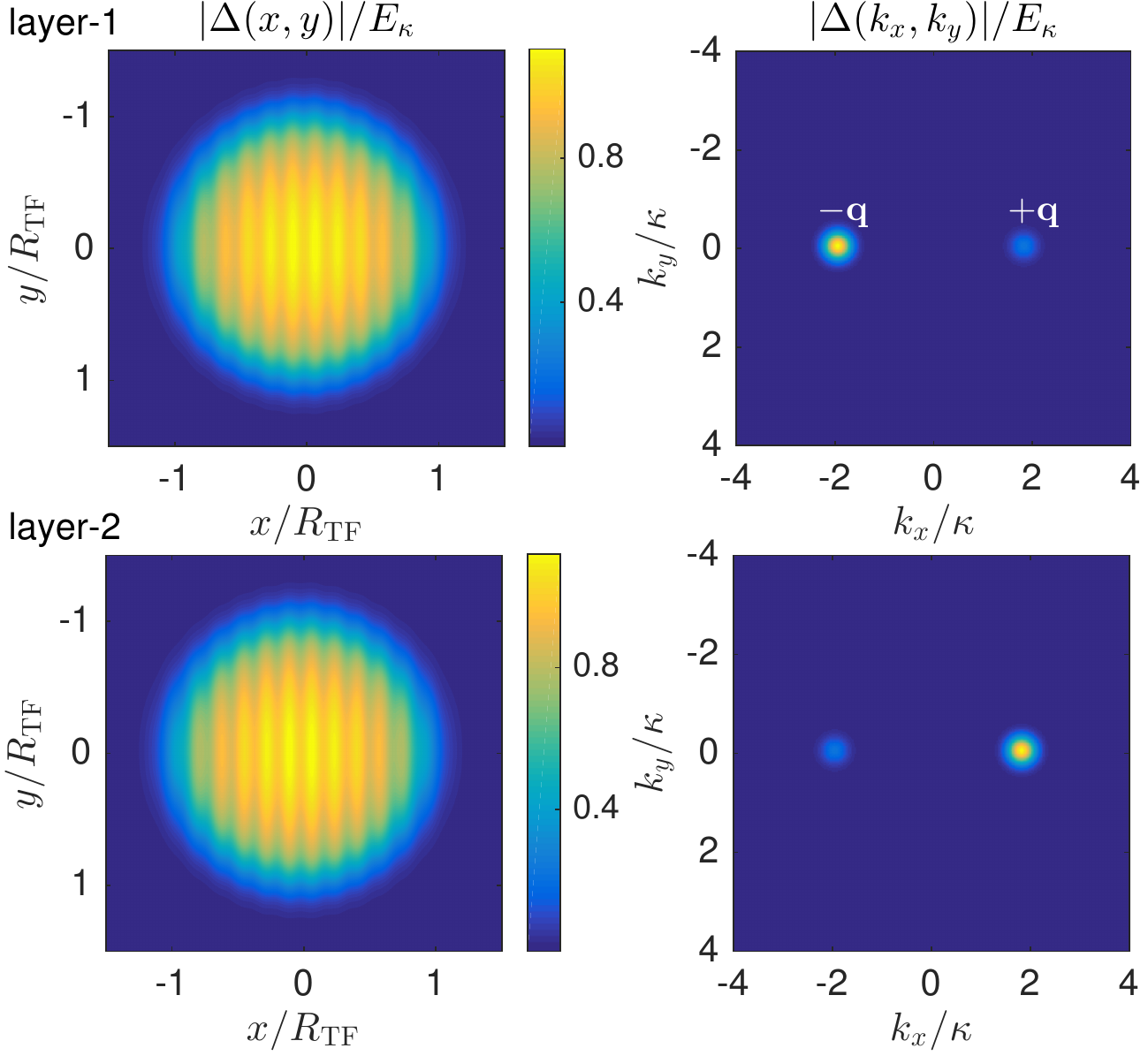}
\caption{{\bf The density profile of the LO order
parameter.} Real-space density profile (left panel) and momentum
space distribution (right panel) of the superfluid order parameters
$\Delta_{1,2}$ for $J/E_\kappa=2$, where two pairing peaks with
opposite momenta $\pm \mathbf{q}$ can be identified in each layer.
In the simulations, we have included a weak harmonic trap with
frequency $\omega\ll E_\kappa$, $R_{\rm
TF}\equiv\sqrt{2E_\kappa/m\omega^2}=\kappa/\omega$,
$E_{b}/E_{\kappa}=1$, and $E_{F}\sim E_{\kappa}$.}
\label{FFLO_state}
\end{figure}

\subsection{Phase Diagram of Superfluidity}
We proceed to study the superfluid phase by computing the real-space superfluid
order parameters
$\Delta_{j}(\mathbf{r})=-U\langle\hat{\psi}_{j,\downarrow}(\mathbf{r})
\hat{\psi}_{j,\uparrow}(\mathbf{r})\rangle$ for the layers $j=1,2$,
and determine the ground state by solving the Bogoliubov-de Gennes
(BdG) equation $H_{{\rm
BdG}}(\mathbf{r})\phi_{\eta}=\varepsilon_{\eta}\phi_{\eta}$, where
$H_{{\rm BdG}}$ is a $8\times8$ matrix, $\phi_{\eta}$ represents the
Nambu basis, and $\varepsilon_{\eta}$ is the corresponding energy of
the Bogoliubov quasiparticles labeled by an index $\eta$ (see Appendix for
the explicit form). The Fourier transformation of the superfluid
order parameters $\Delta_{j}(\mathbf{r})=\sum_\mathbf{q}
\Delta_\mathbf{q} e^{i\mathbf{q}\cdot\mathbf{r}}$ yields different
situations. When $\Delta_\mathbf{q} \neq 0$  for $\mathbf{q} \neq
0$, the system represents a finite-momentum superfluid.  When
$\Delta_\mathbf{q}\neq0$ for  $\mathbf{q}= 0$, the system is in a
conventional superfluid phase.  Otherwise, the system is in a normal
state.

Figure \ref{phase_diagram} illustrates a behavior of the order
parameter $|\Delta_{1,2}|$ and the pairing momentum $|\mathbf{q}|$
as functions of the tunneling strength $J$. One can see that by
increasing $J$, the pairing momentum $|\mathbf{q}|$ decreases
gradually and eventually vanishes at a critical value of the
tunneling strength $\sim4.3E_{\kappa}$. This defines a transition
from the  finite-momentum pairing state to the conventional
superfluid characterized by a zero pairing momentum. Across the
transition, the chemical potential $\mu$ decreases continuously, as
shown in the inset of Fig. \ref{phase_diagram}.

In Fig. \ref{FFLO_state}, we present a density profile of the order
parameter at a moderate tunneling $J/E_{\kappa}=2$. One can find
that the system is in a superfluid phase with the order parameter
$\Delta_{1,2}$ exhibiting an  obvious density-modulation in the
real-space profile (left panel of Fig. \ref{FFLO_state}). The
corresponding momentum distributions of $|\Delta_{1,2}|$ are shown
in the right panel of Fig. \ref{FFLO_state}. Interestingly,  a pair
of peaks  with opposite momenta
$\pm\mathbf{q}\sim\pm1.9\kappa\mathbf{e}_x$ can be identified in
each layer,  so the pairing order parameter contains both momentum
components $\Delta_j\sim c_{j}
e^{-i\mathbf{q}\cdot\mathbf{x}}+d_{j}e^{i\mathbf{q}\cdot\mathbf{x}}$,
with  $c_{j}$ and $d_{j}$ being the corresponding amplitudes. This
results in the spatial modulation of the order parameter with the
periodicity $2\pi/|\mathbf{q}|$. In this way, the LO superfluid
state  is formed which breaks the translational symmetry and
exhibits supersolid properties. Note that the density-modulation of
 the pairing order parameter is gauge invariant. In fact,
 transforming back to the laboratory frame ($\tilde{\psi}_{\gamma1}=e^{i\kappa
x}\psi_{\gamma1}$ and $\tilde{\psi}_{\gamma2}=e^{-i\kappa
x}\psi_{\gamma2}$), the order parameter
$\tilde{\Delta}_j\equiv-U\langle\tilde{\psi}_{j\downarrow}\tilde{\psi}_{j\uparrow}\rangle=
e^{\pm i2\kappa x}\Delta_j$  acquires only a phase factor $e^{\pm
i2\kappa x}$, so the  density modulation is unchanged.

The broad parameter regime shown for the LO phase reflects the
essential difference between the present pseudospin SOC system and
the real SOC Fermi gases, which can be understood in the following
way. In a 1D SOC system  the spin-up and spin-down pockets of the
dispersion are  shifted with respect to each other due to the
momentum transfer between spin-up and spin-down states induced by
the Raman coupling. The Raman coupling also opens a gap in the band
crossing between spin-up and spin-down pockets at zero
momentum~\cite{Liu2009PRL,Lin2011Nature}. In the presence of an
attractive $s$-wave interaction, the favored superfluid pairing
occurs between spin-up and spin-down states with opposite momenta,
respectively in the left-well and right-well pockets of the Fermi
surface, leading to the conventional BCS state. The LO state can be
realized only when the pairing within each (left-well/right-well)
pocket is dominated. This can be in principle achieved by
considering  an attractive $p$-wave interaction which, however, is
currently a major challenge in  cold atom experiments.

\begin{figure}[t]
\centering
\includegraphics[width=0.45\textwidth]{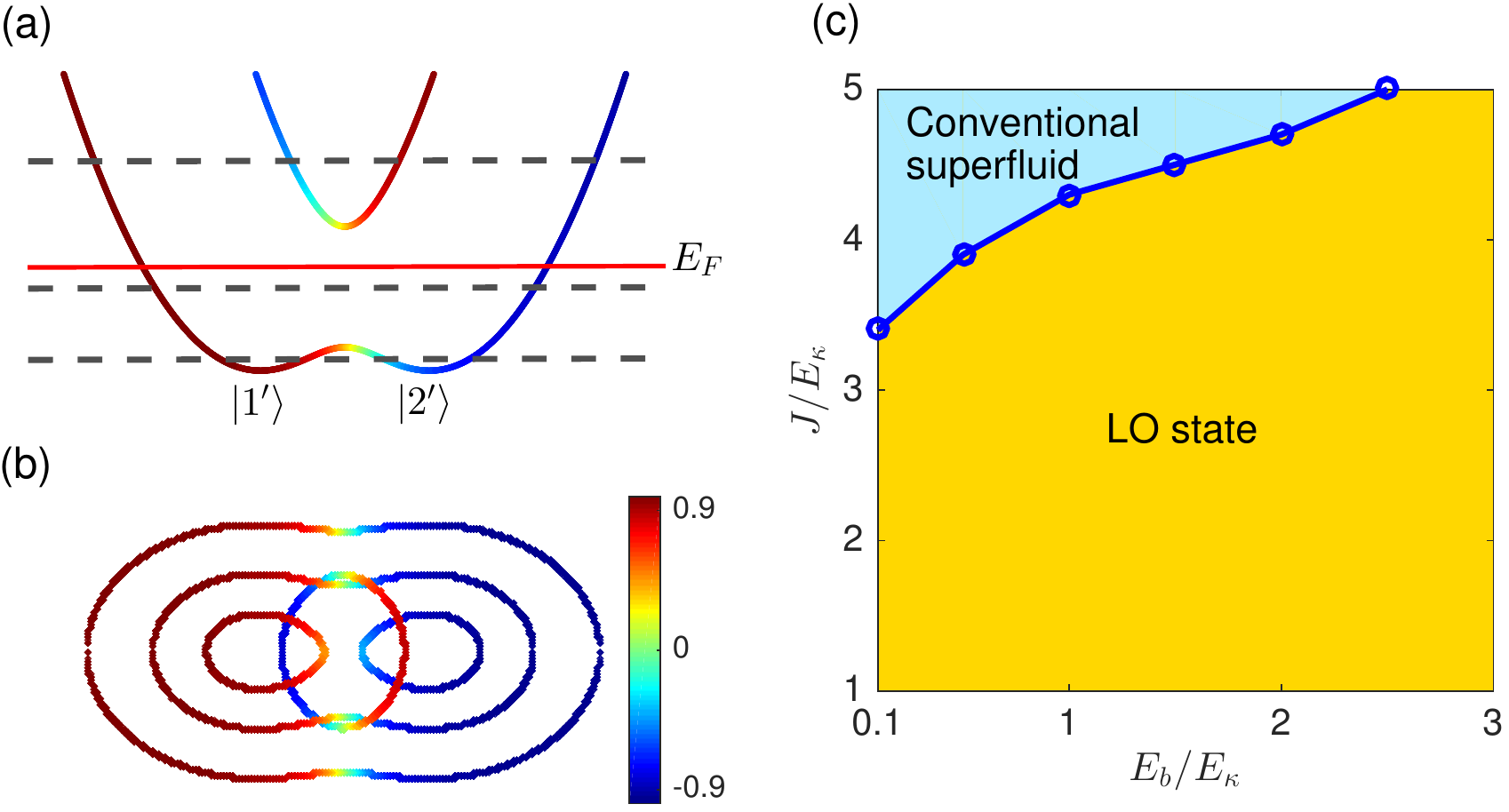}
\caption{{\bf The phase diagram.} (a) Single-particle
spectrum of the bilayer system for $J/E_\kappa=2$. The red solid
line denotes the value of $E_F$ ($\sim E_\kappa$) in the
simulations. (b) Topology of the Fermi surface when the Fermi level
is situated in the lower dispersion branch, the gap, and the upper
branch. The color bar represents the pseudospin polarizations
$P_{\mathbf{k}}=\frac{\langle \hat{n}_{1\mathbf{k}}\rangle-\langle
\hat{n}_{2\mathbf{k}}\rangle}{\langle
\hat{n}_{1\mathbf{k}}\rangle+\langle \hat{n}_{2\mathbf{k}}\rangle}$.
 (c) Phase diagram of superfluidity in the
$J-E_b$ plane with $E_F\sim E_\kappa$. The LO phase is obtained in a very broad parameter regime.} \label{mechanism}
\end{figure}

In comparison, in the present pseudospin SOC system,  the left and
right pockets of the double-well dispersion shown in Fig.
\ref{mechanism}(a,b) correspond to predominantly opposite pseudospin
(i.e. layer) states $|1^\prime\rangle$ and $|2^\prime\rangle$, while
the interlayer tunneling can mix the pseudospin-up and -down states
[Fig. \ref{mechanism}(b)]. From the above analysis one can soon
realize that the LO phase may be realized if there is an attractive
$p$-wave like interaction in the pseudospin space, which induces
pairings within each pocket of the Fermi surface. As explained
below, such an effective $p$-wave like interaction arises naturally
in the present bilayer system with the intralayer $s$-wave
interaction.

Considering the intralayer atomic interaction  $\sim U\int
d\mathbf{r}\sum_j
\hat{n}_{j\downarrow}(\mathbf{r})\hat{n}_{j\uparrow}(\mathbf{r})$,
let us treat the layer states ($j=1,2$) as  fictitious spin states:
 $j\rightarrow s=\Uparrow,\Downarrow$. In turn,
we treat the real spin states ($\gamma=\uparrow,\downarrow$) as an
artificial spatial degree of freedom (e.g. $s$ and $p$ orbital
states) denoted by $A$ and $B$. With these notations, the intralayer
atomic interaction $\sim U\int
d\mathbf{r}\sum_{s=\Uparrow,\Downarrow}
\hat{n}_{s,A}(\mathbf{r})\hat{n}_{s,B}(\mathbf{r})$ describes an
effective  $p$-wave interaction between  the atoms with different
``orbits" $A$ and $B$  and the same ``spin" $s$.  This is
essentially the underlying mechanism for the realization of the LO
phase  in a very broad parameter range, as one can see in
Fig.~\ref{mechanism}c  which shows the phase diagram in the $J-E_b$
plane. Note that the very broad LO phase has been obtained for all
different Fermi energies $E_F$ depicted with dashed lines in
Fig.~\ref{mechanism}a.

The BCS-BEC crossover can be achieved by tuning the binding energy
$E_b$ for the present bilayer pseudospin SOC system. This is
different from the Fermi gases with a 2D Rashba-type SOC in real
spin space, where the BCS-BEC crossover can be achieved by tuning
the Raman recoil energy across the regime that $E_F\sim E_\kappa$,
with the BEC of rashbons being obtained when
$E_\kappa>E_F$~\cite{Hu2011PRL,Yu2011PRL,Shenoy2013JPB}. Note that
$E_b$ is closely related to the 3D $s$-wave scattering length via
$E_b=\frac{\mathcal{C}\omega_z}{\pi}e^{\sqrt{2}\pi l_z/a_s}$, with
$\omega_z$ and $l_z$ being the frequency and length of the axial
trapping perpendicular to the 2D plane \cite{Petrov}.  In the BCS
regime  one has $E_b/E_F\ll1$, and the opposite holds in the BEC
regime with $E_b/E_F\gg 1$. In Fig. \ref{mechanism}c we take
$E_\kappa\sim E_F$ so that $E_b/E_\kappa\sim E_b/E_F$, and the
BCS-BEC crossover with spatial modulations is clearly obtained.

The present results reveal a profound connection between the LO
state of fermions and the stripe phase of bosons. In the BEC regime
with large $E_b$ limit, where $E_b/E_F\gg 1$, all atoms get paired
to tight dimers in each layer. In this situation, the effective
tunneling $\tilde{J}\sim J^2 /\epsilon_B$ of the dimers with the
momentum shift $\pm \tilde{\kappa}\sim \pm 2\kappa$ can induce the
interference of the dimers between the two layers, rendering a
stripe-type phase of the dimer BECs, where $\epsilon_B$ is a binding
energy of the dimer state in the presence of pseudospin SOC. On the
other hand, in the BCS regime with weak pairing condition, the atoms
are loosely paired on the Fermi surface, and the fermionic nature of
the atoms plays the important role. The momentum shift in the
tunneling of individual atoms dominates the spatial modulation of
the effective $p$-wave pairing order, leading to formation of the
fermionic supersolidity of LO state and the superfluid phase
transition, as shown in Fig. \ref{mechanism}c.

\section{Discussion and Conclusion}
In the real experiment, the bilayer geometry for ultracold atoms can be implemented using a double-well superlattice potential which forms a
stack of weakly coupled bilayer systems and has already realized in
experiment for bosons~\cite{Li2017Nature}. The multiple bilayer
configuration can enhance the superfluid transition \cite {Orso} and
increase the signal-to-noise ratio. Note that the typical pairing
momentum $|\mathbf{q}|$ is on the order of the laser recoil
$\kappa$. The corresponding spatial modulation of the LO state can
be readily detected by the optical Brag scattering
\cite{Li2017Nature}.

In conclusion, we have proposed a highly feasible way to realize the
LO phase in the bilayer Fermi gases with both the  TR and SI
symmetries. The laser-assisted tunneling between layers, which
generates a 1D pseudospin SOC, is applied to induce the relative
momentum shift between pairing orders in the bilayer Fermi gas,
leading to the spontaneous formation of LO phase. The realization of
the LO phase is also rooted in a novel mechanism that the intralayer
$s$-wave attractive interaction combined with the pseudospin SOC
renders an effective $p$-wave pairing phase in the pseudospin space.
With the system preserving both TR and SI symmetries, the LO phase
has been predicted in a very broad parameter regime. This work paves
the way to realize the LO phase within the current experimental
accessibility.

\section{acknowledgement}
We thank Joachim Brand, Wolfgang Ketterle, Ana Maria Rey, and Gora
Shlyannikov for helpful discussions and useful suggestions. This
work is supported by the NSFC under Grants No. 11875195, No. 11404225, No.
11474205, and No. 11504037. Q. Sun, A.-C. Ji and X.-J Liu also acknowledge the support by 
the foundation of Beijing Education Committees under Grants  No. CIT\&TCD201804074,
No. KM201510028005, No. KZ201810028043, and the support from the National Key R\&D Program of China (2016YFA0301604), NSFC (No. 11574008 and No. 11761161003), and by the Strategic Priority
Research Program of Chinese Academy of Science (Grant No. XDB28000000).

\appendix
\section{Derivation of the Hamiltonian}
\begin{figure}[t]
\centering \includegraphics[width=0.25\textwidth]{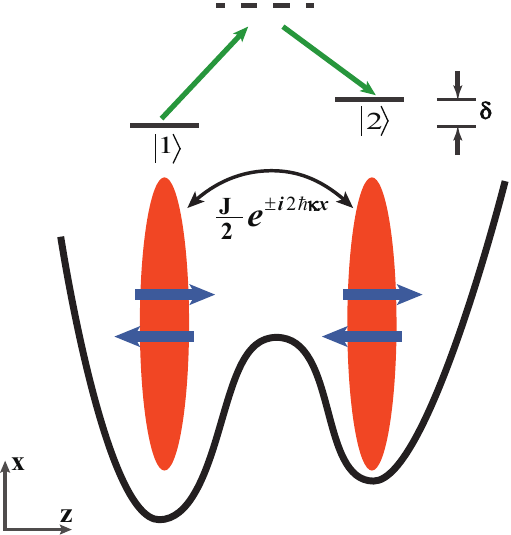}
\caption{Illustrative sketch of the system. A
two-component Fermi gas composed of atoms in two metastable internal
(spin) states are confined in a state-independent double-well
optical potential along the $z$-axis. An asymmetry of the
double-well potential prevents a direct atomic tunneling between the
wells, and instead two Raman lasers are applied to induce a
laser-assisted interlayer tunneling.  Here, $J$ is a strength of the
laser-assisted interlayer tunneling with $2\hbar\kappa$ being an
associated recoil momentum along the $x$-direction,  $\delta$
denotes the energy mismatch between the two layer states.}
\label{Sketch}
\end{figure}

We consider a two-component Fermi gas composed of atoms in two
metastable internal (quasi-spin) states labeled by the index
$\gamma=\uparrow,\downarrow$,  as shown in Fig. \ref{Sketch}. The
atoms are confined in a state-independent double-well optical
potential along the $z$-axis providing a bilayer structure. As in
ref. \cite{Li2016PRL}, we assume an asymmetry of the double-well
potential preventing a direct atomic tunneling between the wells,
and instead two Raman lasers are applied to induce a laser-assisted
interlayer tunneling. In the laboratory frame, the single particle
Hamiltonian for each component is
\widetext
\begin{eqnarray}
H_\gamma=\int
d^2\mathbf{r}\left\{\sum_{j=1,2}\tilde{\psi}^\dag_{\gamma
j}(\mathbf{r}) \left
[\frac{\mathbf{P}^2}{2m}+(-1)^j\frac{h}{2}\pm\frac{\delta}{2}\right
]\tilde{\psi}_{\gamma j}(\mathbf{r}) +\left
(\frac{J}{2}e^{i2\hbar\kappa
\mathbf{e}_x\cdot\mathbf{r}}\tilde{\psi}^\dag_{\gamma 1}(\mathbf{r})
\tilde{\psi}_{\gamma 2}(\mathbf{r})+h.c.\right )\right\},
\label{laboratory}
\end{eqnarray}
\endwidetext
where the upper and lower signs correspond to different atomic
internal states $\gamma$, and $\mathbf{P}$ and $m$ are the momentum
and mass of the atom respectively. $\tilde{\psi}_{\gamma
j}(\mathbf{r})$ annihilates a Fermi atom at position $\mathbf{r}$ in
the $j$-th layer with an internal state $\gamma$. $J$ is a strength
of the laser-assisted interlayer tunneling with $2\hbar\kappa$ being
an associated recoil momentum along the $x$-direction. $h$ and
$\delta$ denote the energy mismatch between the two layers and two
components. Note that, both spin components are characterized by the
same laser-assisted tunneling.  In writing Eq.(\ref{laboratory}) we
have neglected the off-resonant transition by applying the Rotating
Wave Approximation (RWA) for the laser-assisted interlayer
tunneling.

In terms of $\tilde{\psi}_{\gamma j}(\mathbf{r})$, the atom-atom coupling described by the contact interaction between the
fermions within individual layers in different internal states $\gamma$ can be written as
\begin{eqnarray}
H_{\rm int}=\frac{U}{2}\sum_{\gamma,j}\int d^2\mathbf{r}\tilde{\psi}^\dag_{\gamma j}(\mathbf{r})\tilde{\psi}_{\gamma j}(\mathbf{r})
\tilde{\psi}^\dag_{\bar{\gamma} j}(\mathbf{r})\tilde{\psi}_{\bar{\gamma} j}(\mathbf{r}).
\label{interaction_lab}
\end{eqnarray}
with $U$ the contact interaction strength in two dimension.

We now apply a gauge transformation $\psi_{\gamma 1}=e^{-i\hbar\kappa\mathbf{e}_x\cdot\mathbf{r}}\tilde{\psi}_{\gamma 1}$
and $\psi_{\gamma 2}=e^{i\hbar\kappa\mathbf{e}_x\cdot\mathbf{r}}\tilde{\psi}_{\gamma 2}$. Then, Hamiltonian (\ref{laboratory}) becomes
\widetext
\begin{eqnarray}
H_\gamma=\int d^2\mathbf{r}\left\{\sum_{j=1,2}\psi^\dag_{\gamma
j}(\mathbf{r})\left [\frac{\mathbf{(P+(-1)^{j+1}
\hbar\kappa}\mathbf{e}_x)^2}{2m}+(-1)^j\frac{h}{2}\pm\frac{\delta}{2}\right
]\psi_{\gamma j}(\mathbf{r})+\frac{J}{2} \left (\psi^\dag_{\gamma
1}(\mathbf{r})\psi_{\gamma 2}(\mathbf{r})+h.c.\right )\right\},
\label{transformed}
\end{eqnarray}
\endwidetext
while the form of the interaction  Hamiltonian, Eq.
(\ref{interaction_lab}), is unchanged.  By making use of the Fourier
transformation $\psi_{\gamma
j}(\mathbf{r})=\frac{1}{\sqrt{S}}\sum_{\mathbf{k}}\psi_{\gamma
j}(\mathbf{k})e^{i\hbar\mathbf{k}\cdot\mathbf{r}}$ and setting
$\hbar$ to the unity, we obtain the  the momentum space Hamiltonian
given by Eq.(1) in the main text.

\section{Two-body bound state}
Let us begin with the two-body problem, which helps to intuitively understand the many-body
behavior of this system. The molecular state with center-of-mass (COM)
momentum $\mathbf{q}$ can be constructed as $|\Phi\rangle_{\mathbf{q}}=\frac{1}{2}\sum_{\mathbf{k}}\sum_{j,l=1}^{2}\phi_{\mathbf{k}
\mathbf{q},jl}S_{\mathbf{k}\mathbf{q},jl}^{\dag}\left|0\right\rangle$ with
\begin{equation}
S_{\mathbf{k}\mathbf{q},jl}^{\dag}=\frac{1}{\sqrt{2}}[\psi_{j\uparrow}^{\dag}(\mathbf{Q}_{+})\psi_{l\downarrow}^{\dag}(\mathbf{Q}_{-})
-\psi_{j\downarrow}^{\dag}(\mathbf{Q}_{+})\psi_{l\uparrow}^{\dag}(\mathbf{Q}_{-})],
\end{equation}
where $\mathbf{Q}_{\pm}=\mathbf{q}/2\pm\mathbf{k}$ and $S_{\mathbf{qk},jl}^{\dag}$ is a singlet operator for creating a pair of atoms in the layers $j$
and $l$ with a COM momentum $\mathbf{q}$ and a relative momentum
$\mathbf{k}$, and $\phi_{\mathbf{k}\mathbf{q},jl}$ is the corresponding probability amplitude. Here, we have used the fact that the four singlet operators introduced above form an invariant subspace of the total Hamiltonian $\mathcal{H}$ and hence serve as a complete basis for the bound state. Notice that due to relations $S_{-\mathbf{k}\mathbf{q},jl}^{\dag}=S_{\mathbf{k}\mathbf{q},lj}^{\dag}$ and $\phi_{-\mathbf{k}\mathbf{q},jl}=\phi_{\mathbf{k}\mathbf{q},lj}$, we have added a factor $1/2$ in the summations  $\sum_{\mathbf{k}}$ to avoid a double counting of the terms comprising the state vector $|\Phi\rangle_{\mathbf{q}}$.

Substituting the wave-function $|\Phi\rangle_{\mathbf{q}}$ into the stationary Schr\"{o}dinger equation $\mathcal{H}|\Phi\rangle_{\mathbf{q}}
=E_{\mathbf{q}}|\Phi\rangle_{\mathbf{q}}$ one can get the following explicit form for each component:
\widetext
\begin{eqnarray}
E_{\mathbf{q}}\phi_{\mathbf{kq},11} & = &
\{\mathbf{k}^{2}+(\mathbf{q}/2+\kappa\mathbf{e}_{x})^{2}-h\}\phi_{\mathbf{kq},11}+\frac{J}{2}(\phi_{\mathbf{kq},12}
+\phi_{\mathbf{kq},21})+\frac{U}{S}\sum_{\mathbf{k}^{\prime}}\phi_{\mathbf{k^\prime q},11}\\
E_{\mathbf{q}}\phi_{\mathbf{kq},12} & = & \{(\mathbf{k}+\kappa\mathbf{e}_{x})^{2}+\mathbf{q}^{2}/4\}\phi_{\mathbf{kq},12}+\frac{J}{2}(\phi_{\mathbf{kq},11}+\phi_{\mathbf{kq},22})\\
E_{\mathbf{q}}\phi_{\mathbf{kq},21} & = & \{(\mathbf{k}-\kappa\mathbf{e}_{x})^{2}+\mathbf{q}^{2}/4\}\phi_{\mathbf{kq},21}+\frac{J}{2}(\phi_{\mathbf{kq},11}+\phi_{\mathbf{kq},22})\\
E_{\mathbf{q}}\phi_{\mathbf{kq},22} & = &
\{\mathbf{k}^{2}+(\mathbf{q}/2-\kappa\mathbf{e}_{x})^{2}+h\}\phi_{\mathbf{kq},22}+\frac{J}{2}(\phi_{\mathbf{kq},12}+\phi_{\mathbf{kq},21})+\frac{U}{S}\sum_{\mathbf{k}^{\prime}}\phi_{\mathbf{k^\prime
q},22}.
\end{eqnarray}
\endwidetext
where $E_{\mathbf{q}}$ is an eigenenergy.
Then, we have
\widetext
\begin{eqnarray}
E_{\mathbf{q}}(\phi_{\mathbf{kq},11}+\phi_{\mathbf{kq},22}) &=& (\mathbf{k}^{2}+\mathbf{q}^{2}/4+\kappa^{2})(\phi_{\mathbf{kq},11}
+\phi_{\mathbf{kq},22})+(q_{x}\kappa-h)(\phi_{\mathbf{kq},11}-\phi_{\mathbf{kq},22})\\
&+& \Omega(\phi_{\mathbf{kq},12}+\phi_{\mathbf{kq},21})+\frac{U}{S}\sum_{\mathbf{k}^{\prime}}(\phi_{\mathbf{k^\prime q},11}+\phi_{\mathbf{k^\prime q},22})\\
E_{\mathbf{q}}(\phi_{\mathbf{kq},11}-\phi_{\mathbf{kq},22}) &=& (\mathbf{k}^{2}+\mathbf{q}^{2}/4+\kappa^{2})(\phi_{\mathbf{kq},11}-\phi_{\mathbf{kq},22})+(q_{x}\kappa-h)(\phi_{\mathbf{kq},11}+\phi_{\mathbf{kq},22})\\
&+& \frac{U}{S}\sum_{\mathbf{k}^{\prime}}(\phi_{\mathbf{k^\prime q},11}-\phi_{\mathbf{k^\prime q},22})\\
(\phi_{\mathbf{kq},12}+\phi_{\mathbf{kq},21}) &=& \left\{
\frac{J/2}{E_{\mathbf{q}}-(\mathbf{k}+\kappa\mathbf{e}_{x})^{2}
-\mathbf{q}^{2}/4}+\frac{J/2}{E_{\mathbf{Q}}-(\mathbf{k}-\kappa\mathbf{e}_{x})^{2}-\mathbf{q}^{2}/4}\right\}
(\phi_{\mathbf{kq},11}+\phi_{\mathbf{kq},22}).
\end{eqnarray}
\endwidetext

Solving the above equations, one finds
\widetext
\begin{eqnarray}
(\phi_{\mathbf{kq},11}+\phi_{\mathbf{kq},22})=\frac{\gamma_{\mathbf{k}}\frac{U}{S}\sum_{\mathbf{k}^{\prime}}(\phi_{\mathbf{k^\prime q},11}+\phi_{\mathbf{k^\prime q},22})+\beta_{\mathbf{k}}\frac{U}{S}\sum_{\mathbf{k}^{\prime}}(\phi_{\mathbf{k}^{\prime},1}-\phi_{\mathbf{k}^{\prime},4})}{\alpha_{\mathbf{k}}\gamma_{\mathbf{k}}-\beta_{\mathbf{k}}^{2}}\\
(\phi_{\mathbf{kq},11}-\phi_{\mathbf{kq},22})=\frac{\alpha_{\mathbf{k}}\frac{U}{S}\sum_{\mathbf{k}^{\prime}}(\phi_{\mathbf{k^\prime q},11}-\phi_{\mathbf{k^\prime q},22})+\beta_{\mathbf{k}}\frac{U}{S}\sum_{\mathbf{k}^{\prime}}(\phi_{\mathbf{k^\prime q},11}+\phi_{\mathbf{k^\prime q},22})}{\alpha_{\mathbf{k}}\gamma_{\mathbf{k}}-\beta_{\mathbf{k}}^{2}},
\end{eqnarray}
\endwidetext
with $\alpha_{\mathbf{k}}=E_{\mathbf{q}}-(\mathbf{k}^{2}+\mathbf{q}^{2}/4+\kappa^{2})$, $\beta_{\mathbf{k}}=q_{x}\kappa-h$, and $\gamma_{\mathbf{k}}=\alpha_{\mathbf{k}}-\left\{
\frac{J^{2}/2}{E_{\mathbf{q}}-(\mathbf{k}+\kappa\mathbf{e}_{x})^{2}
-\mathbf{q}^{2}/4}+\frac{J^{2}/2}{E_{\mathbf{q}}-(\mathbf{k}-\kappa\mathbf{e}_{x})^{2}-\mathbf{q}^{2}/4}\right\}$.


After some straightforward derivations, we arrive at the following self-consistent equation
for the molecular energy $E_{\mathbf{q}}$:
\widetext
\begin{eqnarray}
(\frac{U}{S}\sum_{\mathbf{k}}\frac{\alpha_{\mathbf{k}}}{\alpha_{\mathbf{k}}\gamma_{\mathbf{k}}-\beta_{\mathbf{k}}^{2}}-1)(\frac{U}{S}\sum_{\mathbf{k}}\frac{\gamma_{\mathbf{k}}}{\alpha_{\mathbf{k}}\gamma_{\mathbf{k}}-\beta_{\mathbf{k}}^{2}}-1)-(\frac{U}{S}\sum_{\mathbf{k}}\frac{\beta_{\mathbf{k}}}{\alpha_{\mathbf{k}}\gamma_{\mathbf{k}}-\beta_{\mathbf{k}}^{2}})^{2}=0.\label{molecular_energy}
\end{eqnarray}
\endwidetext
Notice that, $\alpha_{\mathbf{k}}$, $\beta_{\mathbf{k}}$ and $\gamma_{\mathbf{k}}$ do not depend on the detuning $\delta$ between two components, i.e. the two-body spectrum obtained using Eq. (\ref{molecular_energy}) would not be altered by changing $\delta$. On the other hand, the detuning shifts the single particle spectrum by $\pm\delta/2$ for different spin components.

Minimizing $E_{\mathbf{q}}$ with respect to the COM momentum $\mathbf{q}$,
one can obtain the ground state energy of the molecular state. The
corresponding coefficients $\phi_{\mathbf{kq},jl}$ (not normalized) are given by
\widetext
\begin{eqnarray}
\sum_{\mathbf{k}^{\prime}}(\phi_{\mathbf{k^\prime q},11}-\phi_{\mathbf{k^\prime q},22}) & = & \sum_{\mathbf{k}^{\prime}}(\phi_{\mathbf{k^\prime q},11}+\phi_{\mathbf{k^\prime q},22})(\frac{U}{S}\sum_{\mathbf{k}}\frac{-\beta_{\mathbf{k}}}{\alpha_{\mathbf{k}}\gamma_{\mathbf{k}}-\beta_{\mathbf{k}}^{2}})/(\frac{U}{S}\sum_{\mathbf{k}}\frac{\alpha_{\mathbf{k}}}{\alpha_{\mathbf{k}}\gamma_{\mathbf{k}}-\beta_{\mathbf{k}}^{2}}-1)\\
(\phi_{\mathbf{kq},11}+\phi_{\mathbf{kq},22}) & = & \frac{\gamma_{\mathbf{k}}\frac{U}{S}\sum_{\mathbf{k}^{\prime}}(\phi_{\mathbf{k^\prime q},11}+\phi_{\mathbf{k^\prime q},22})+\beta_{\mathbf{k}}\frac{U}{S}\sum_{\mathbf{k}^{\prime}}(\phi_{\mathbf{k^\prime q},11}-\phi_{\mathbf{k^\prime q},22})}{\alpha_{\mathbf{k}}\gamma_{\mathbf{k}}-\beta_{\mathbf{k}}^{2}}\\
(\phi_{\mathbf{kq},11}-\phi_{\mathbf{kq},22}) & = & \frac{\alpha_{\mathbf{k}}\frac{U}{S}\sum_{\mathbf{k}^{\prime}}(\phi_{\mathbf{k^\prime q},11}-\phi_{\mathbf{k^\prime q},22})+\beta_{\mathbf{k}}\frac{U}{S}\sum_{\mathbf{k}^{\prime}}(\phi_{\mathbf{k^\prime q},11}+\phi_{\mathbf{k^\prime q},22})}{\alpha_{\mathbf{k}}\gamma_{\mathbf{k}}-\beta_{\mathbf{k}}^{2}}\\
\phi_{\mathbf{kq},12} & = & \frac{J/2}{E_{\mathbf{q}}-(\mathbf{k}+\kappa\mathbf{e}_{x})^{2}-\mathbf{q}^{2}/4}(\phi_{\mathbf{kq},11}+\phi_{\mathbf{kq},22})\\
\phi_{\mathbf{kq},21} & = &
\frac{J/2}{E_{\mathbf{q}}-(\mathbf{k}-\kappa\mathbf{e}_{x})^{2}-\mathbf{q}^{2}/4}(\phi_{\mathbf{kq},11}+\phi_{\mathbf{kq},22}).
\end{eqnarray}
\endwidetext
For vanishing interlayer and intercomponent detunings, i.e.
$h=\delta=0$, one recovers the results in the main text. Some
remarks: (i) For zero recoil $\kappa=0$, we have
$\beta_{\mathbf{k}}=0$ and Eq. (\ref{molecular_energy}) reduces to
$\frac{U}{S}\sum_{\mathbf{k}}\frac{1}{\alpha_{\mathbf{k}}}-1=0$ and
$\frac{U}{S}\sum_{\mathbf{k}}\frac{1}{\gamma_{\mathbf{k}}}-1=0$,
with the binding energy $E_{b}^{(1)}=\sqrt{E_{b}^{2}+J^{2}}-J$ and
$E_{b}^{(2)}=E_{b}-J$ modified simply by the tunneling strength $J$.
Compared with usual $E_{b}$, we see that the binding energies are
modified simply by the tunneling strength $J$. (ii) For $J=0$, the
momentum transfer $2\kappa\mathbf{e}_{x}$ brought by the Raman
coupling can be simply gauged away via a unitary transformation
describing the layer-dependent momentum shift, and the binding
energy is $E_{b}^{(1)}=E_{b}^{(2)}=E_{b}$. (iii) For $\kappa\neq0$
and $J\neq0$, the attractive interaction between atoms in different
internal states would act together with the interlayer tunneling and
intra-component spin-orbit coupling. This can give rise to
nontrivial two-body and many-body ground states discussed in the
main text.

\section{Bogoliubov-de Gennes equation of the system}
Since the atomic attraction takes place only in the same layer, we introduce the superfluid order parameters $\Delta_j(\textbf{r})=-U\left<\psi_{j,\downarrow}(\mathbf{r})
\psi_{j,\uparrow}(\mathbf{r})\right>$ ($j=1,2$) with $\mathbf{r}=(x,y)$. Then, the Hamiltonian (1) in the main text can be diagonalized via a Bogoliubov--Valatin transformation. By taking into account an additional weak harmonic trapping potential $V(r)=m\omega^2r^2/2$, the resultant Bogoliubov-de Gennes (BdG) equation can be written as: $H_{\rm BdG}(\mathbf{r})\phi_{\eta}=\varepsilon_{\eta}\phi_{\eta}$.
Here,
\begin{equation}
H_{\mathrm{BdG}}(\mathbf{r})=\left( \begin{matrix}H_{1}(\mathbf{r}) & H_{J}\\
H^\dagger_{J} & H_{2}(\mathbf{r})\end{matrix}\right)\, \label{H_BdG}
\end{equation} is an $8\times 8$ matrix with $H_{J}=\mathrm{diag}(J/2,J/2,-J/2,-J/2)$ describing the interlayer tunneling, and $H_{1,2}(\mathbf{r})$ denoting the single-particle Hamiltonian for each layer $j=1,2$. The latter $H_{j}(\mathbf{r})$ reads explicitly
\begin{equation}
\begin{split}
H_{j}(\mathbf{r}) &=
\begin{pmatrix}
\epsilon_{j\uparrow}(\mathbf{r}) & 0 & 0 & -\Delta_{j}(\mathbf{r})\\
0 & \epsilon_{j\downarrow}(\mathbf{r}) & \Delta_{j}(\mathbf{r}) & 0\\
0 & \Delta_{j}^{\ast}(\mathbf{r}) & -\epsilon_{j\uparrow}^{\ast}(\mathbf{r}) & 0\\
-\Delta_{j}^{\ast}(\mathbf{r}) & 0 & 0 & -\epsilon_{j\downarrow}^{\ast}(\mathbf{r})\\
\end{pmatrix}\,,\\
\end{split}
\end{equation}
with $j=1,2$ and
\begin{equation}
\left\{
\begin{aligned}
\epsilon_{1\uparrow,\downarrow}(\mathbf{r})&=-\hbar^{2}\nabla^2/(2m)-i\hbar^{2}\kappa\partial_{x}/m+V(\mathbf{r})-\mu\\
\epsilon_{2\uparrow,\downarrow}(\mathbf{r})&=-\hbar^{2}\nabla^2/(2m)+i\hbar^{2}\kappa\partial_{x}/m+V(\mathbf{r})-\mu\\
\end{aligned}
\right.\,.
\end{equation}
Here, we have taken $h=\delta=0$. In this case, the ground state is spin balanced with equal chemical potential $\mu$ for both components. The Nambu basis is chosen as $\phi_{\eta}=[u_{1\uparrow,\eta},u_{1\downarrow,\eta},
v_{1\uparrow,\eta},v_{1\downarrow,\eta},u_{2\uparrow,\eta},u_{2\downarrow,\eta},
v_{2\uparrow,\eta},v_{2\downarrow,\eta}]^{T}$, and $\varepsilon_{\eta}$ is the corresponding energy of  the Bogoliubov quasiparticles labeled by an index $\eta$.
The order parameter $\Delta_{1,2}(\mathbf{r})$ is to be determined self-consistently by
\begin{equation}
\Delta_{j}(\mathbf{r})=-U\sum_{\eta}[u_{j\uparrow,\eta}
v^{\ast}_{j\downarrow,\eta}f(-\varepsilon_{\eta})\nonumber+u_{j\downarrow,\eta}v^{\ast}_{j\uparrow,\eta}f(\varepsilon_{\eta})]\,,
\end{equation}
where $f(E)=1/[e^{E/k_{B}T}+1]$ is the Fermi-Dirac distribution function
at a temperature $T$. The chemical potential $\mu$ is obtained using
the number equation $N=\int d\mathbf{r} n(\mathbf{r})$, where the
total atomic density is given by
\begin{equation}
n(\mathbf{r})=\sum_{j\gamma,\eta}[|u_{j\gamma,\eta}(\mathbf{r})|^{2}
f(\varepsilon_{\eta})+|v_{j\gamma,\eta}(\mathbf{r})|^{2}f(-\varepsilon_{\eta})]\,.
\end{equation}
The ground state can then be found by solving the above BdG equation self-consistently with the basis expansion method \cite{Xu2014PRL}. In the numerical simulations, we have taken a large energy cutoff
$\varepsilon_{c}=6E_{\mathrm{rec}}$ to ensure the accuracy of the calculation, where
$E_{\mathrm{rec}}=5\hbar\omega$ assures that the trap oscillation frequency $\omega$ is much smaller than the recoil frequency.

\end{document}